\documentclass[11pt]{article}
\usepackage[dvips]{graphics,graphicx}
\DeclareGraphicsExtensions{.eps}

\usepackage{a4}
\def\ds{\displaystyle}

\def\nn{\nonumber}
\def\re{\mbox{\sf Re}}
\def\im{\mbox{\sf Im}}

\def\bspace{\!\!\!\!\!\!\!\!\!\!}

\def\qed{\hbox{${\vcenter{\vbox{                        
   \hrule height 0.4pt\hbox{\vrule width 0.4pt height 6pt
   \kern5pt\vrule width 0.4pt}\hrule height 0.4pt}}}$}}

\def\R{\bf\sf R}
\def\Z{\bf\sf Z}
\def\C{\bf\sf C}

\newtheorem{theorem}{Theorem}
\newtheorem{lemma}{Lemma}
\newtheorem{conjecture}{Conjecture}

\newtheorem{definition}{Definition}

\begin{document}

\begin{center}
  
  {\bf \Large Fractals of the Julia and Mandelbrot sets of the Riemann
    Zeta Function}

\bigskip

{\large S.C. Woon}

\medskip

Trinity College, University of Cambridge, Cambridge CB2 1TQ, UK

s.c.woon@damtp.cam.ac.uk

MSC-class Primary 28A80, 11S40, 11P32


December 27, 1998

\bigskip

{\bf \large Abstract}

\end{center}

Computations of the Julia and Mandelbrot sets of the Riemann zeta
function and observations of their properties are made.  In the
appendix section, a corollary of Voronin's theorem is derived and a
scale-invariant equation for the bounds in Goldbach conjecture is
conjectured.

\section{Introduction and Motivation}

In this paper, we shall present a series of diagrams showing the Julia
set of the Riemann zeta function and its related Mandelbrot set.  This
computation of the Julia and Mandelbrot sets of the Riemann zeta
function may be, with reference to the existing literature, the first
attempt ever made.

As we shall see, the Julia and Mandelbrot sets of the Riemann zeta
function have unique features and are quite unlike those of any
elementary functions.

We shall also point out, in the Appendix, two known observations of
approximate self-similarity in Number Theory --- in the images of
Riemann zeta function itself, and in the problem of the Goldbach
conjecture where a functional equation or scale-invariant equation for
the ds in Goldbach conjecture is conjectured.

\section{Properties of the Iterated Maps of the Riemann Zeta Function}

\begin{definition}[{\bf Iterated Maps of the Riemann zeta function}]\quad\\
  Denote the iterated maps of the Riemann zeta function $\zeta(s)$ as
  $\,\zeta_n : \C \,\mapsto\, \C \;$ such that
\begin{equation}
\zeta_n : s \;\mapsto\; \underbrace{\zeta(\zeta(\cdots\zeta}_{n 
\mbox{\footnotesize-times}}(s)\cdots))
\end{equation}
\end{definition}

\begin{definition}[{\bf Attractor and Repellor}]\quad\\
  A fixed point $p$ of $f(x)$ is an attractor if $|f'(p)| < 1$, a
  repellor if $|f'(p)| > 1$.
\end{definition}

\begin{lemma}[{\bf An Asymptotic Property of $\zeta(s)$}]\quad\\
\label{l:zeta_asymp}
\begin{equation}
\log |\zeta(-|\sigma|)| = |\sigma| \log(|\sigma|\!+\!1) - 
|\sigma| (1\!+\!\log 2\pi) + O(\log |\sigma|) \quad (\sigma\in\R)\;.
\end{equation}
\end{lemma}

{\bf Proof}

From the functional equation of the Riemann zeta function
\begin{equation}
\zeta(1\!-\!s) = 2 \;(2\pi)^{-s} \;\Gamma(s)\; \cos\left(\frac{\pi
    s}{2}\right) \;\zeta(s)\;,
\end{equation}
we have
$$|\zeta(\sigma)| \le |(2\pi)^\sigma \pi^{-1} \Gamma(1\!-\!\sigma)
\zeta(1\!-\!\sigma)|$$
as $|\sin(\sigma \pi/2)| \le 1$.

As $\sigma \to \infty$,
$\zeta(\sigma) \to 1$ faster than $(1 + \sigma^{-q}) \to 1$ for any
positive real $q$,
\begin{eqnarray}
\zeta(1\!+\!|\sigma|) &=& 1 + O(|\sigma|^{-q})\;,\nn\\
\log(\zeta(1\!+\!|\sigma|) &=& O(|\sigma|^{-q})\;,\nn\\
\log(\Gamma(1\!+\!|\sigma|)) &=& (1+|\sigma|) \log(1+|\sigma|) -
(1+|\sigma|) + O(\log|\sigma|)\nn\\
&&\qquad \mbox{ (by Stirling's formula) }.\nn
\end{eqnarray}
Thus,
\begin{eqnarray}
\log |\zeta(-|\sigma|)|
 &=& -|\sigma| \log(2\pi) + (1+|\sigma|) \log(1+|\sigma|) -
 (1+|\sigma|) + O(\log|\sigma|)\nn\\
 &=& |\sigma| \log(|\sigma|+1) - |\sigma| (1+\log 2\pi) +
 O(\log|\sigma|)\;.\nn
\end{eqnarray}
\hfill\qed

\begin{theorem}[{\bf Fixed Points of $\zeta_n$}]\quad\\
\label{t:fixed_points}
The Riemann zeta function $\zeta(s)$ has one attractor fixed point at
$$\sigma = \alpha = -0.2959050055752\ldots$$
but infinitely many repellor fixed points.
\end{theorem}

{\bf Proof}

For $\sigma>0$, $y=\sigma$ intersects $y=\zeta(\sigma)$ at only one
point, $\sigma = \alpha_+ = 1.833772\dots$. $\zeta(\alpha_+)$ is thus
a fixed point of $\zeta_n$.  Since $|\zeta'(\alpha_+)| = 1.3742\ldots
> 1$, the fixed point at $\sigma = \alpha_+$ is a repellor.

For $-1<\sigma<0$, $y=\sigma$ intersects $y=\zeta(\sigma)$ at $\sigma
= \alpha = -0.2959050055752\ldots$.  $\zeta(\alpha)$ is thus a fixed
point of $\zeta_n$.  Since $|\zeta'(\alpha)| = 0.51273\ldots < 1$, the
fixed point at $\sigma = \alpha$ is an attractor.

For $\sigma \le -1$, $\zeta(2\sigma\!+\!1) = 0$ at $\sigma \in \Z^-$,
and $\zeta(\sigma)$ oscillates with amplitude $\gg |\sigma|$ (by Lemma
\ref{l:zeta_asymp}).  Thus for $\sigma \le -1$, $y=\sigma$ intersects
$y = \zeta(\sigma)$ twice for the interval of every two zeros of
$\zeta(\sigma)$, and $|\zeta'(\sigma)| \gg 1$ for $\sigma$ in the
neighbourhood of these intersections.  $\zeta_n$ has thus infinitely
many repellor fixed points for $\sigma \le -1$.
\hfill\qed

\section{Julia set and Mandelbrot set of $\zeta(s)+a$ for $s,a\in\C$}

For the quadratic map $F(a,s) = s^2 + a$ where $s,a\in\C$, we have the
following definitions.

\begin{definition}[{\bf Julia set}]\quad\\
  The Julia set $J$ of a map $F$ is the boundary of the domains of the
  attractors of the iteration of $F$.
\end{definition}

For example, if $A_1,\; A_2$ are attractors of the iteration of a map
$F$, $D(A_1),\; D(A_2)$ are the domains of the attractors
respectively, and $\partial D(A_1),\; \partial D(A_2)$ are the
boundaries of the domains respectively, then the Julia set of $F$ is
$J = \partial D(A_1) = \partial D(A_2)$.

\begin{definition}[{\bf Mandelbrot set}]\quad\\
  The Mandelbrot set $M$ of a map $F(a,s)$, for a chosen initial
  iteration value $s=s_0$, is the set of values of the complex
  parameter $a$ which, in the iteration of $F(a,s)$ with the chosen
  initial value $s_0$, does not lie in the domain of attraction of
  complex infinity.
\begin{equation}
M = \{a : a\in\C, \lim_{n\to\infty} F_n(a,s_0) \not\to \infty\}\;.
\end{equation}
\end{definition}

The Mandelbrot set is related to the Julia set in the following theorem.

\begin{theorem}[{\bf The Julia-Fatou Theorem}]
  \cite[p. 105]{nonlinear_systems}\\
\label{t:JuliaFatou}
  The Julia set $J$ of a map $F(a,s)$ for the given value of $a$ is
  connected if and only if ${\ds\lim_{n\to\infty}} F_n(a,0)
  \not\to\infty$, i.e. if and only if $a$ belongs to the Mandelbrot
  set $M$ of the map $F(a,s)$ for $s=0$.
\end{theorem}

Though the above definitions are for polynomials and are not the most
general for non-polynomials, we shall choose to adopt these
definitions here to study, for starters, a particular one-parameter
family of the iterated maps of the Riemann zeta function, i.e., that
of $F(a,s) = \zeta(s) + a$.

\bigskip

The Riemann zeta function $\zeta(s)$ has one attractor at fixed point
$s = \alpha$ (by Theorem \ref{t:fixed_points}).  It is not known if
$\zeta(s)$ has any non-real fixed point, and this remains to be proved.

\bigskip

The Julia set of $\zeta(s)$ has been numerically computed with {\em
  Mathematica} with the algorithm downloadable from the {\em Internet}
  \ at

  http://www.damtp.cam.ac.uk/user/scw21/riemann/

\begin{figure}[hbt]
\vskip .5truein
\begin{center}
\begin{tabular}{rl}
$\bspace\bspace$\raisebox{2.5truein}{$\im(s)$}&
\includegraphics{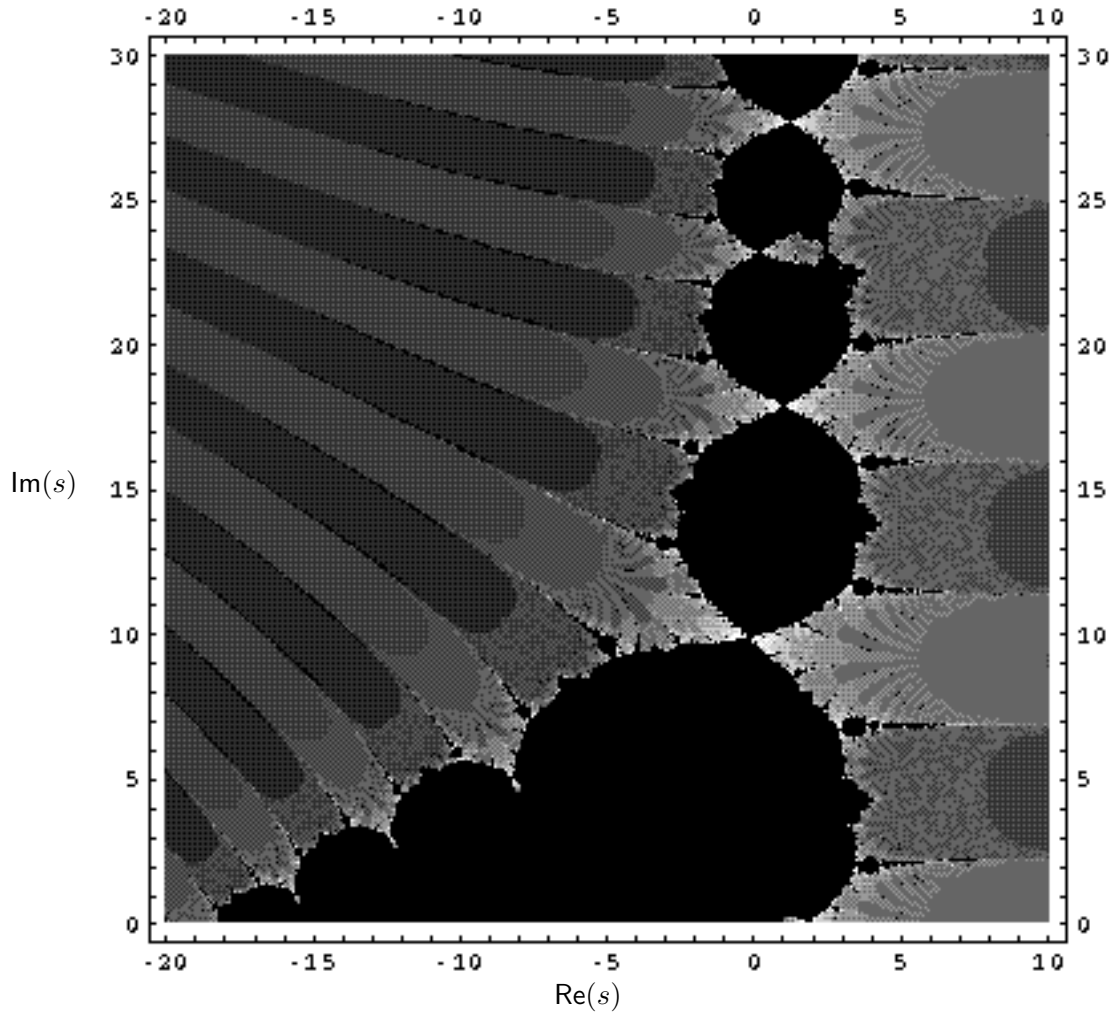}
\end{tabular}
$\re(s)$
\caption{The Julia set of $\zeta(s)$ for $-20\le\re(s)\le 10\;\;$ and 
$\;\;0\le\im(s)\le 30\,$.}
\label{fig:zjulia_1}
\end{center}
\end{figure}

\begin{figure}[hbt]
\vskip .5truein
\begin{center}
\begin{tabular}{rl}
$\bspace\bspace$\raisebox{2.5truein}{$\im(s)$}&
\includegraphics{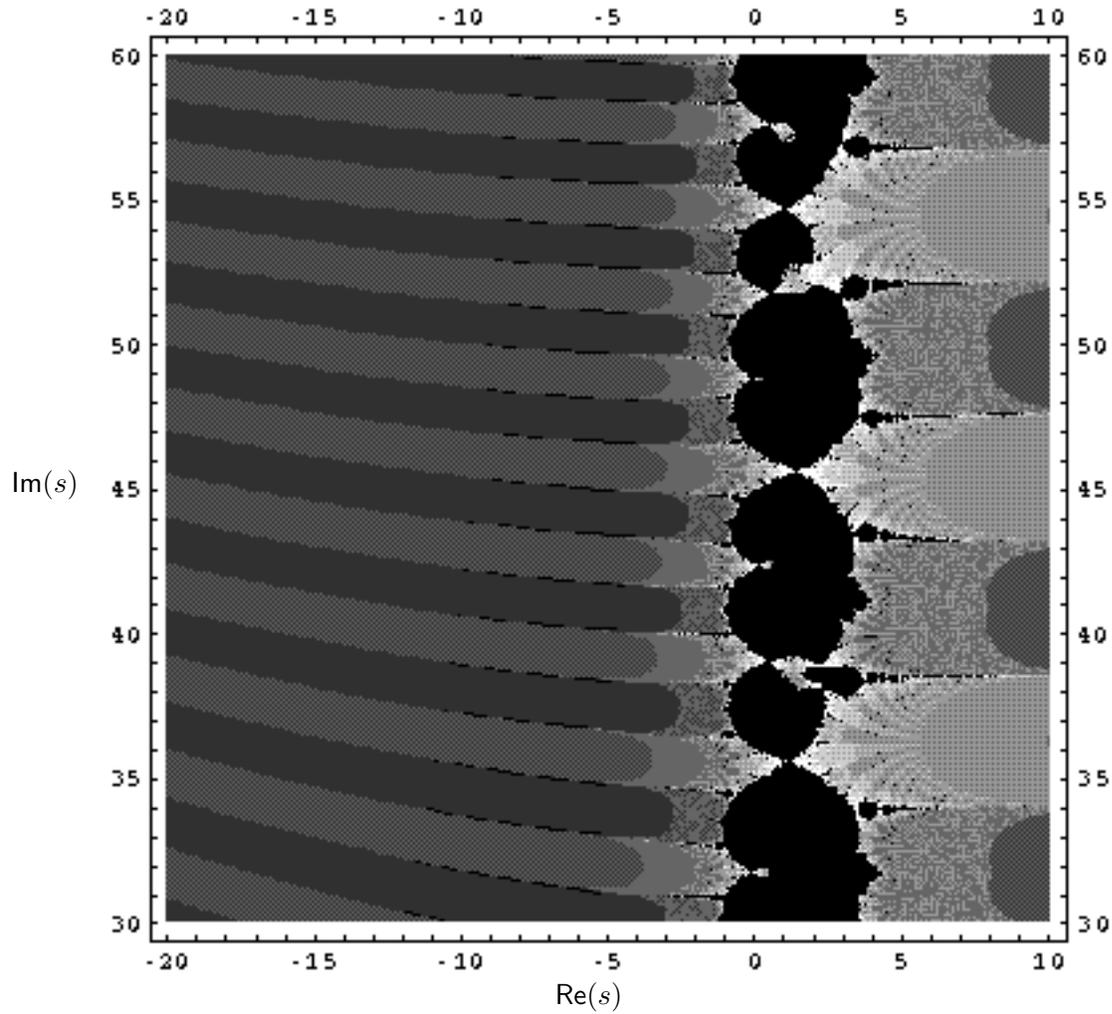}
\end{tabular}
$\re(s)$
\caption{The Julia set of $\zeta(s)$ for $-20\le\re(s)\le 10\;\;$ and 
$\;\;30\le\im(s)\le 60\,$.}
\label{fig:zjulia_2}
\end{center}
\end{figure}

\begin{figure}[hbt]
\vskip .5truein
\begin{center}
\begin{tabular}{rl}
$\bspace\bspace$\raisebox{2.5truein}{$\im(s)$}&
\includegraphics{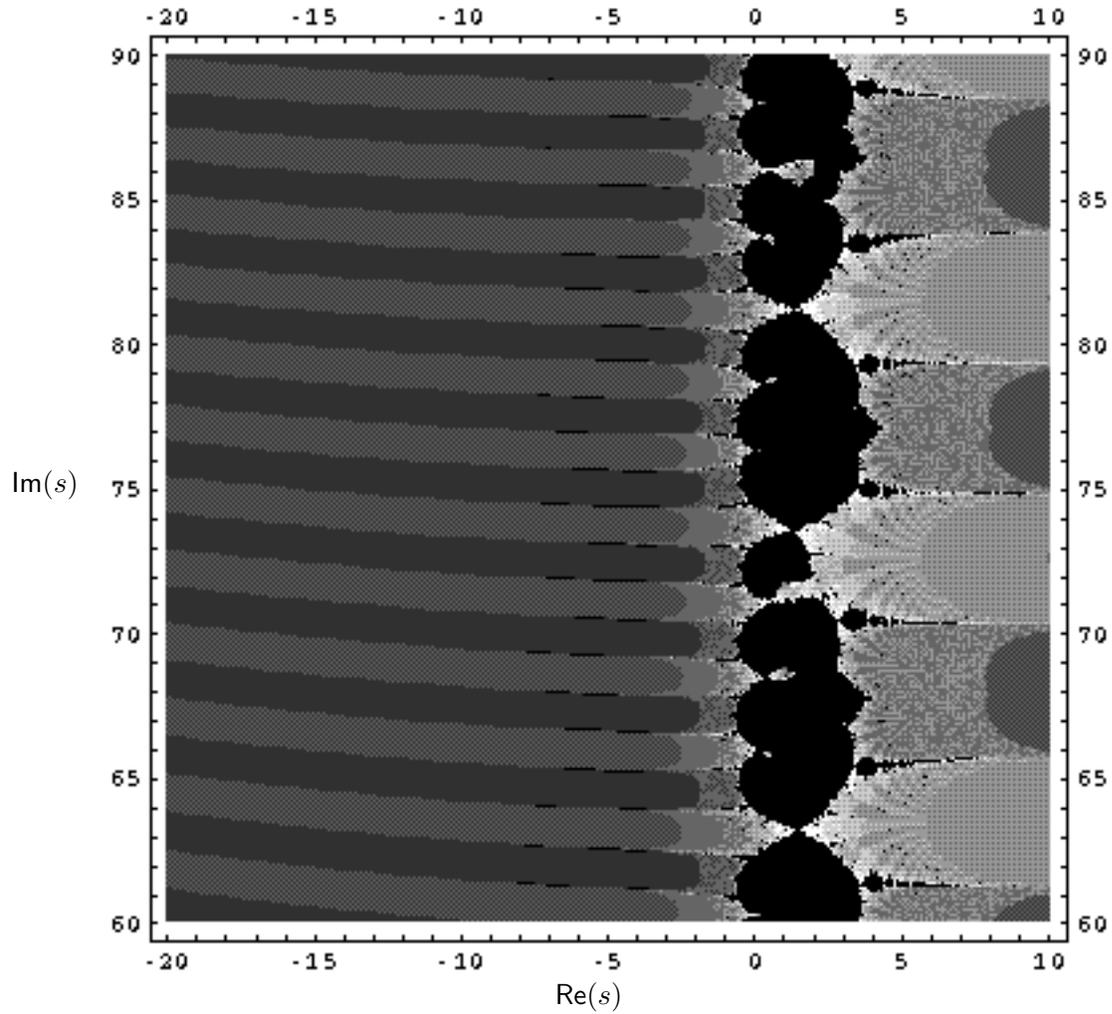}
\end{tabular}
$\re(s)$
\caption{The Julia set of $\zeta(s)$ for $-20\le\re(s)\le 10\;\;$ and 
$\;\;60\le\im(s)\le 90\,$.}
\label{fig:zjulia_3}
\end{center}
\end{figure}

\begin{figure}[hbt]
\vskip .5truein
\begin{center}
\begin{tabular}{rl}
$\bspace\bspace$\raisebox{2.5truein}{$\im(s)$}&
\includegraphics{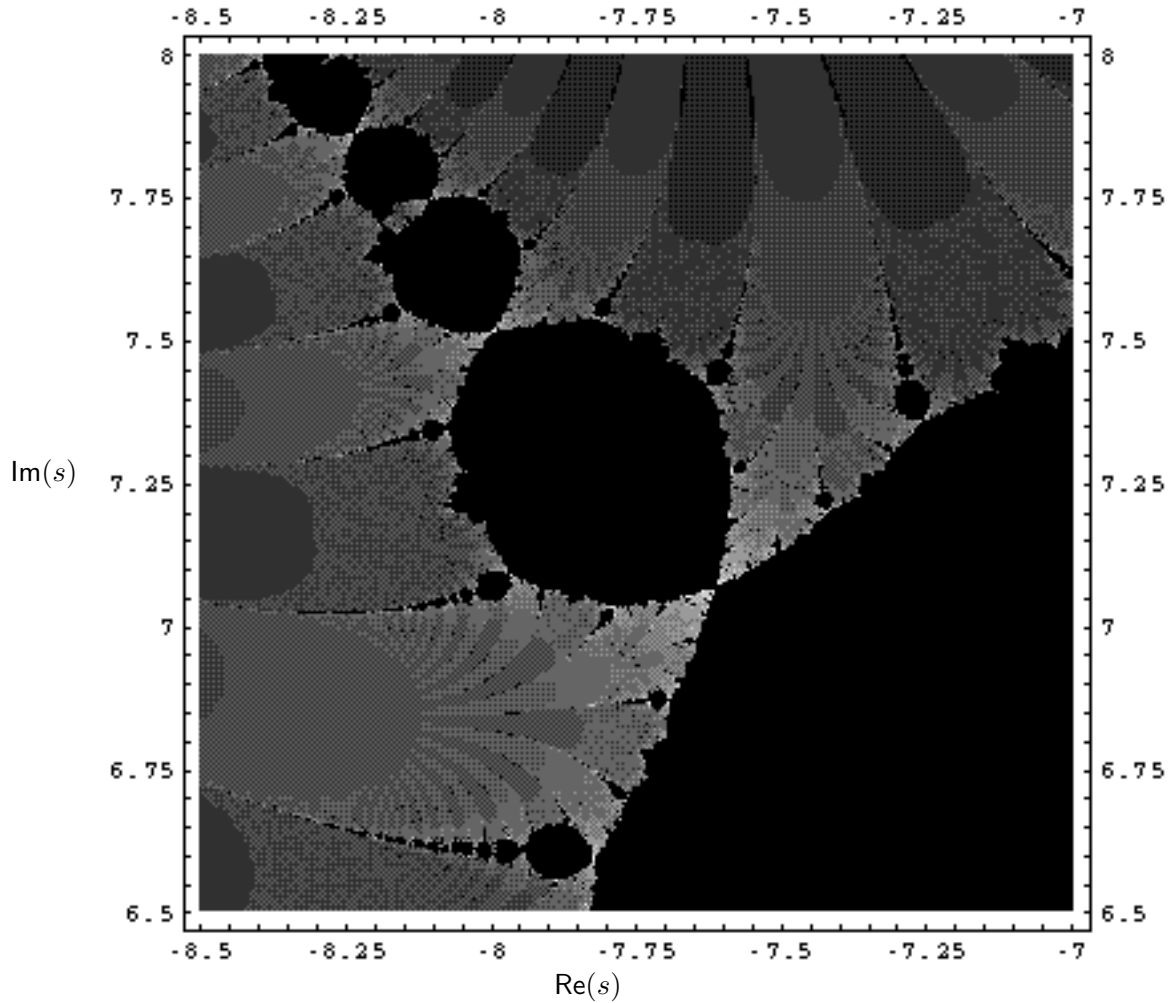}
\end{tabular}
$\re(s)$
\caption{Zooming in to reveal the self-similarity of the 
Julia  set of $\zeta(s)$ for 
$-8.5\le\re(s)\le -7\;\;$ and $\;\;6.5\le\im(s)\le 8\,$.}
\label{fig:zjulia_zoom}
\end{center}
\end{figure}

The results are presented in Figures \ref{fig:zjulia_1},
\ref{fig:zjulia_2}, \ref{fig:zjulia_3} and \ref{fig:zjulia_zoom}.
This may be the very first computation of the Julia set of $\zeta(s)$
ever attempted.

\newpage

The black region represents the basins of attraction of the attractor
of the fixed point at $s=\alpha$ and those of attracting cycles.  The
other attractor is at $\infty$.  The Julia set is the boundary of the
black region.

The regions of different shades represent the various rates of
amplitudes in iteration $\to\infty$, with the lighter shades having
slower rates, and the darker shades, faster rates.

There is a reflection symmetry about the $\re(s)$-axis due to the
complex conjugate property of an analytic complex function.

\newpage

Although a ``piece'' of the black region up along the $\im(s)$-axis
may resemble some other nearby ``piece'', when examined in detail,
each ``piece''is unique and is distinct from any other ``piece''.

\newpage

The ``pieces'' of the black region seem to run indefinitely up the
$\im(s)$-axis and lies on $\re(s)>0$ thereon.

\newpage

\begin{figure}[hbt]
\begin{center}
\includegraphics[height=1.65truein]{zjulia1.eps}\\
(i) $a=0$

\begin{tabular}{cc}
\includegraphics[height=1.65truein]{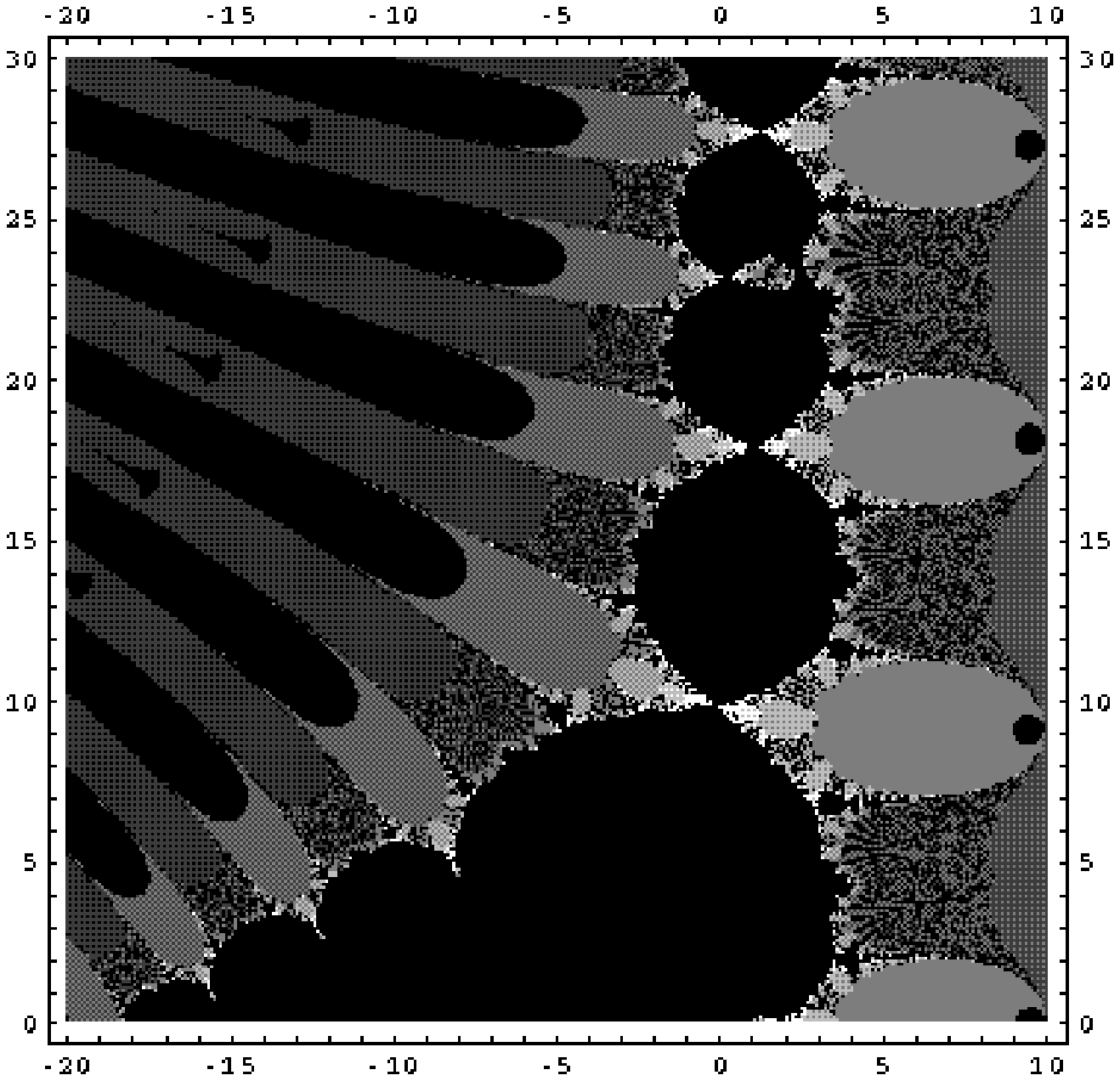}&
\includegraphics[height=1.65truein]{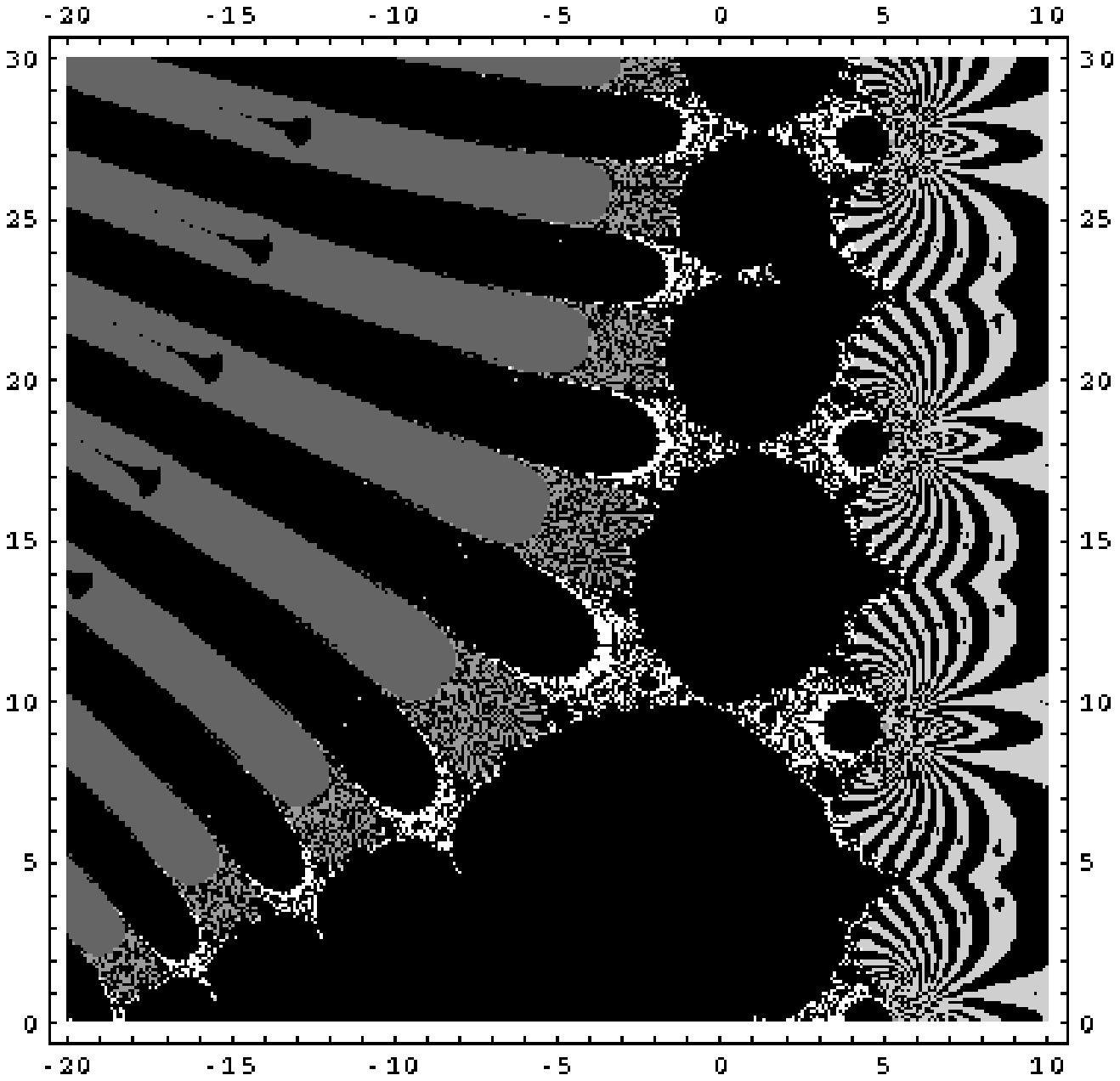}\\
(ii) $a=-0.001$&(iii) $a=-0.01$\\
\includegraphics[height=1.65truein]{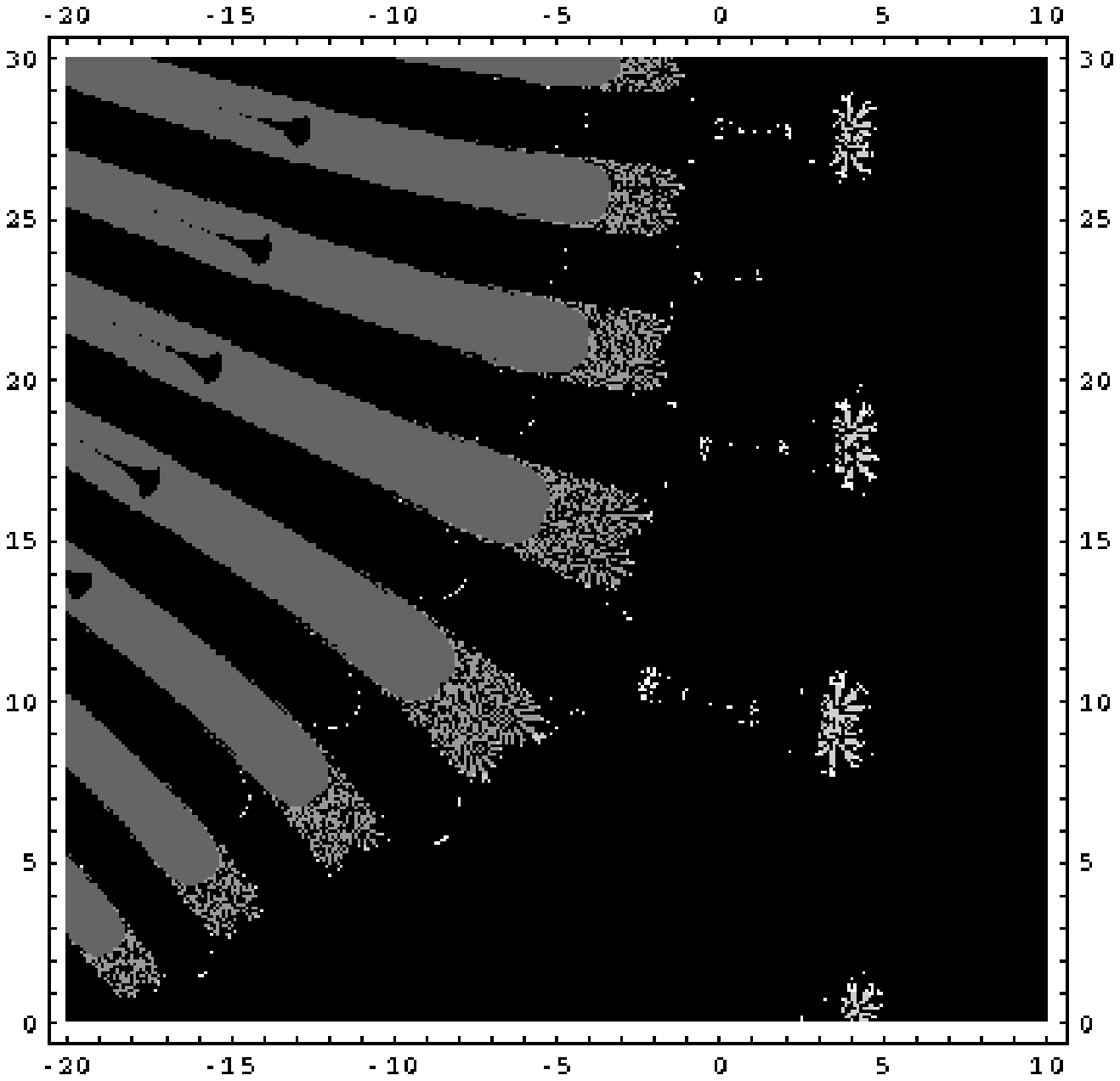}&
\includegraphics[height=1.65truein]{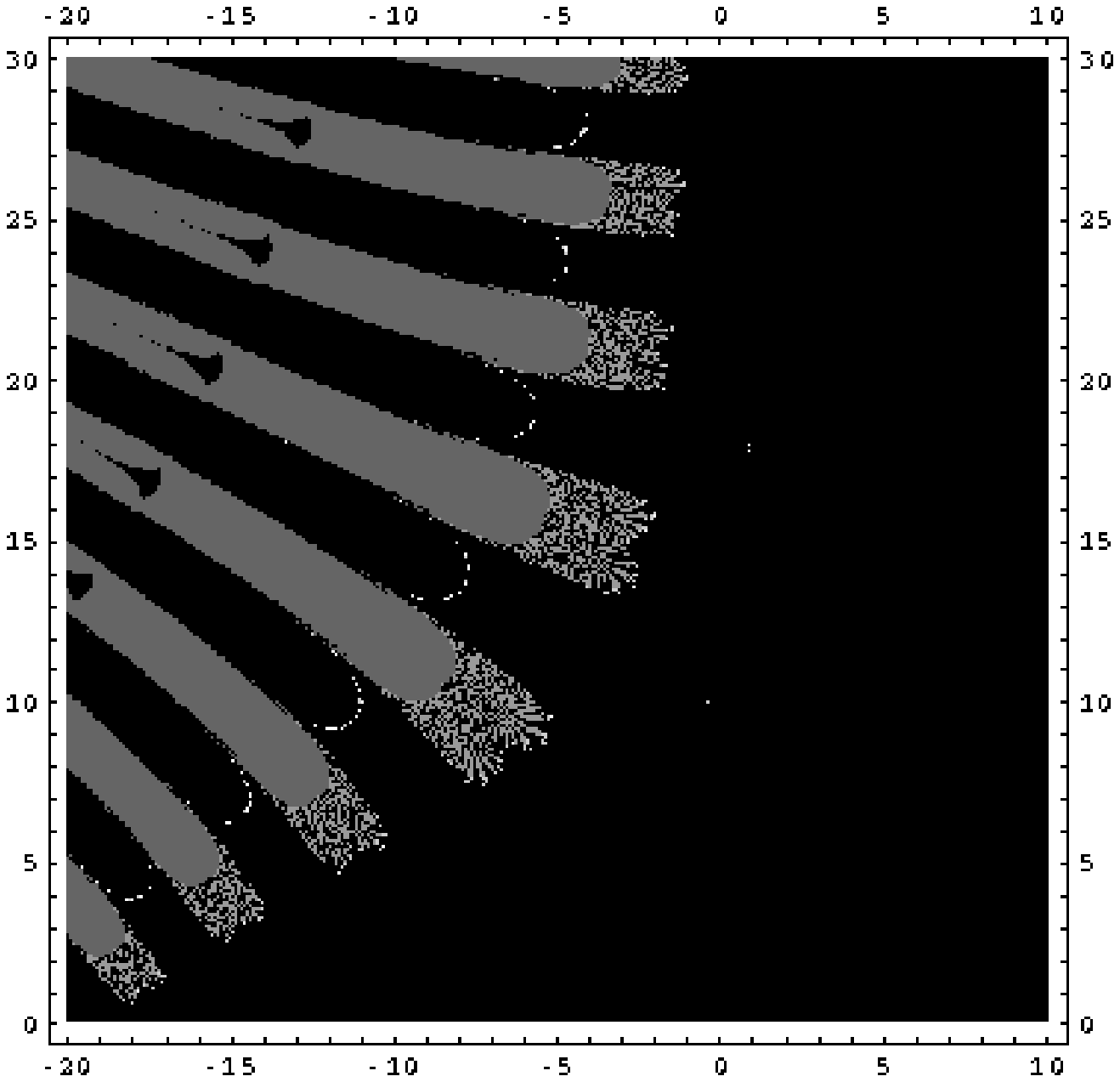}\\
(iv) $a=-0.1$&(v) $a=-1$
\end{tabular}
\caption{Julia sets of $F(a,s) = \zeta(s) + a$}
for $-20\le\re(s)\le 10\,$, $0\le\im(s)\le 30\,$, and\\
(i) $a=0$, (ii) $a=-0.001$, (iii) $a=-0.01$, 
(iv) $a=-0.1$, (v) $a=-1$.\\
The horizontal axis is $\re(s)$, and the vertical axis is $\im(s)$.
\label{fig:zjulia_evolve-}
\end{center}
\end{figure}

\begin{figure}[hbt]
\begin{center}
\includegraphics[height=1.65truein]{zjulia1.eps}\\
(i) $a=0$

\begin{tabular}{cc}
\includegraphics[height=1.65truein]{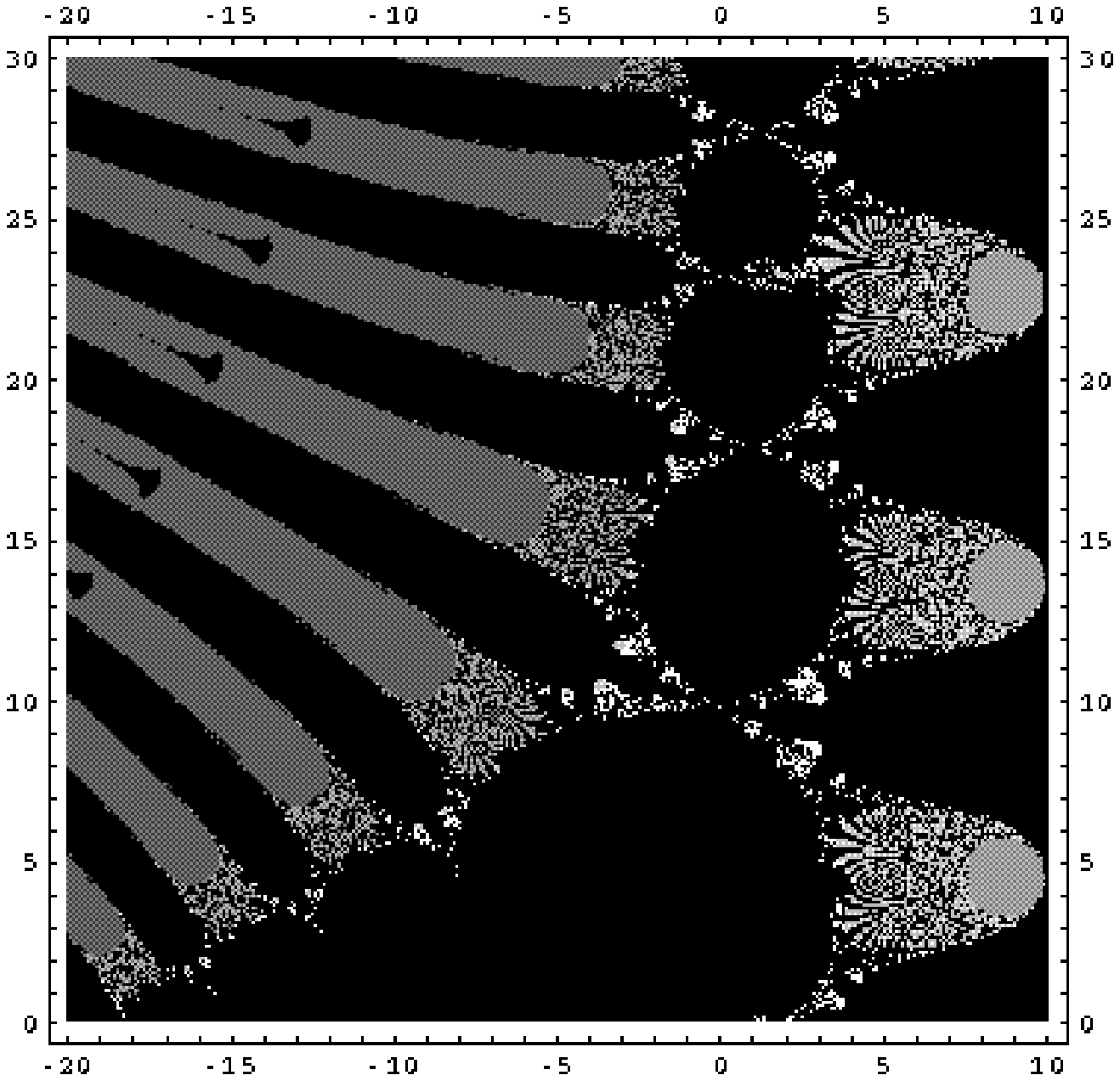}&
\includegraphics[height=1.65truein]{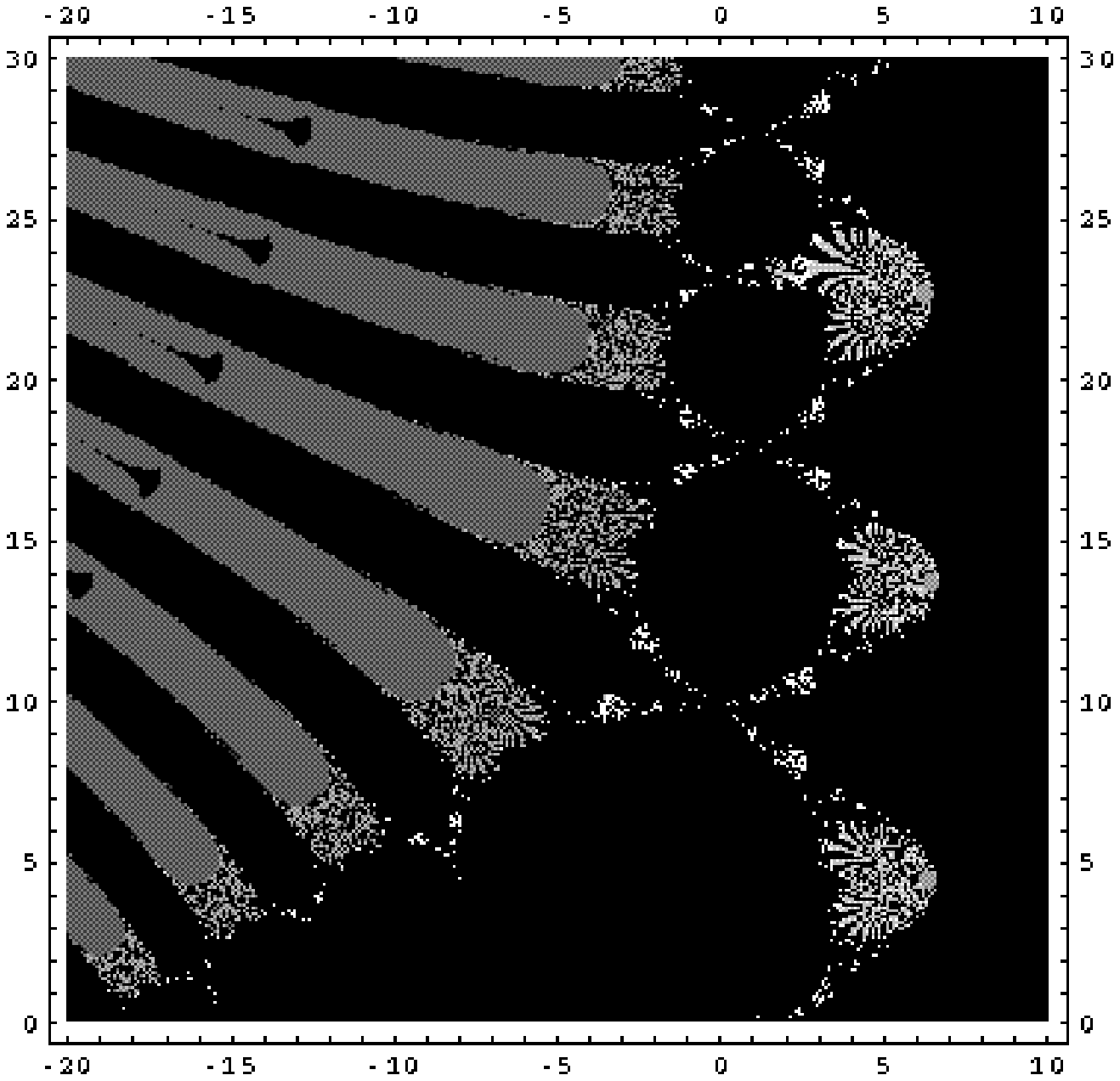}\\
(ii) $a=0.001$&(iii) $a=0.01$\\
\includegraphics[height=1.65truein]{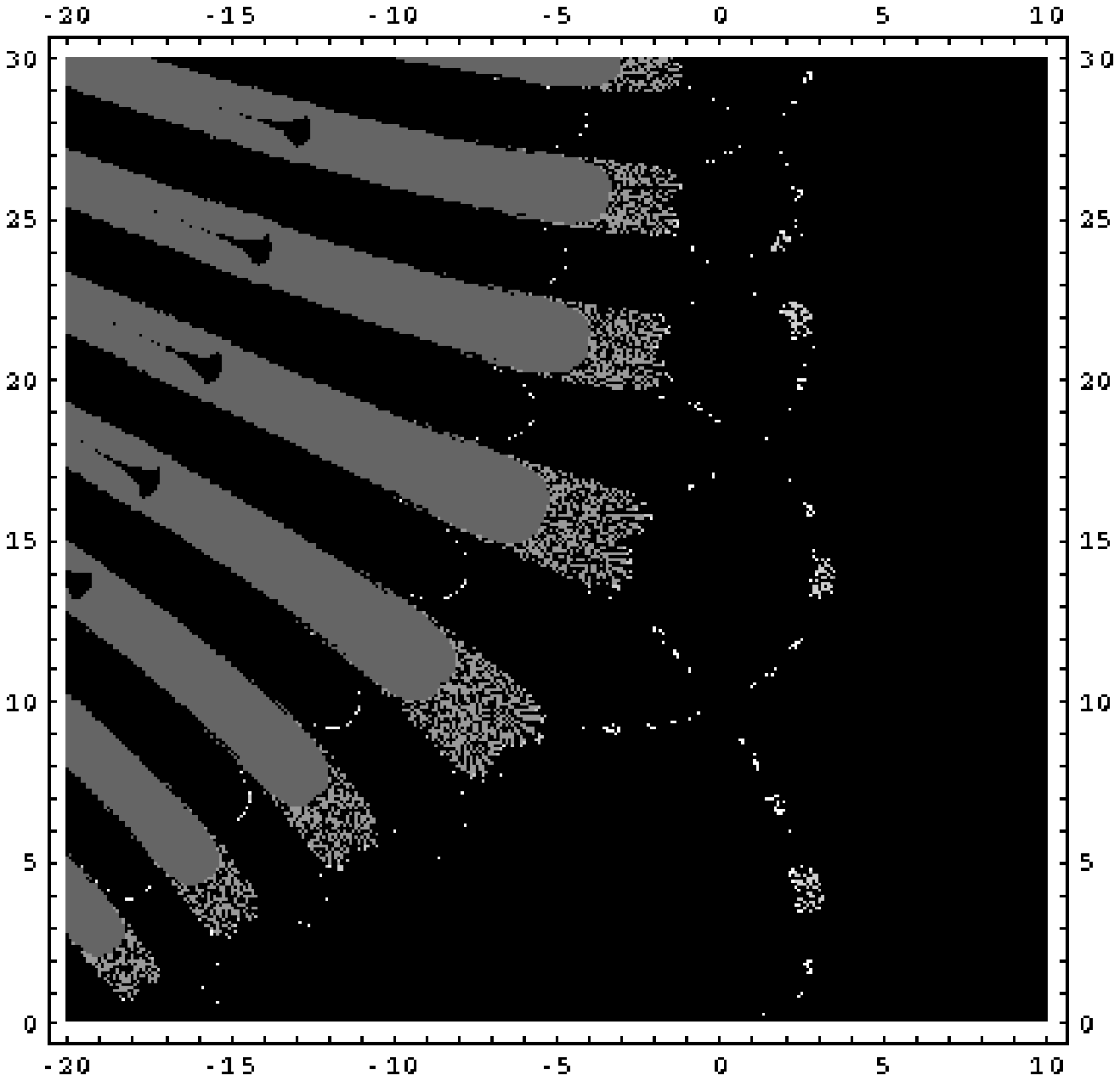}&
\includegraphics[height=1.65truein]{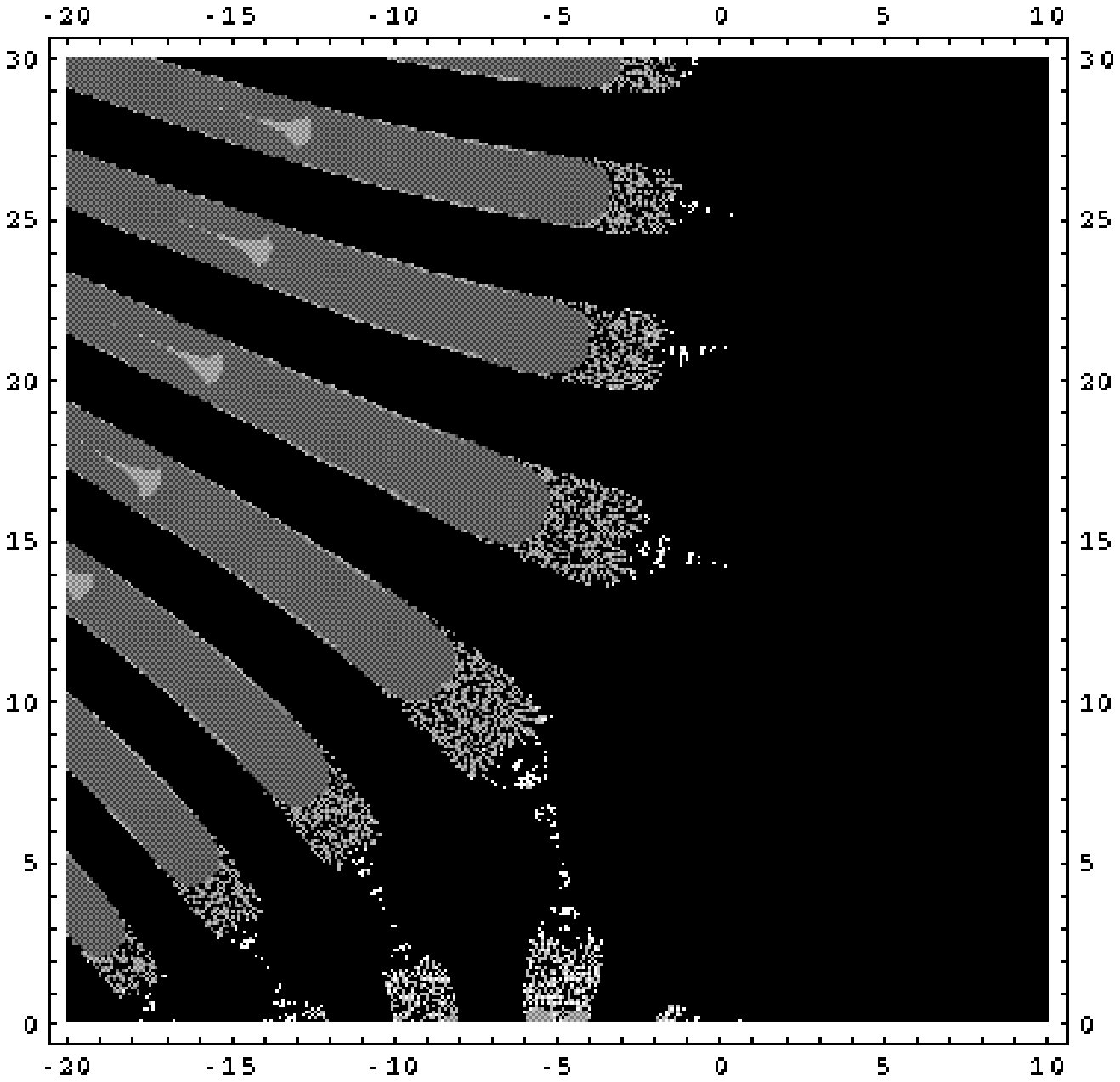}\\
(iv) $a=0.1$&(v) $a=1$
\end{tabular}
\caption{Julia sets of $F(a,s) = \zeta(s) + a$}
for $-20\le\re(s)\le 10\,$, $0\le\im(s)\le 30\,$, and\\
(i) $a=0$, (ii) $a=0.001$, (iii) $a=0.01$, 
(iv) $a=0.1$, (v) $a=1$.\\
The horizontal axis is $\re(s)$, and the vertical axis is $\im(s)$.
\label{fig:zjulia_evolve+}
\end{center}
\end{figure}

We shall now turn to the Julia sets of $F(a,s) = \zeta(s) + a$ for
$a\in\C$.  We shall evolve the Julia sets of $F(a,s)$ from one value
of $a$ to another along a path in the complex plane of $a$, and study
the dynamics in the change of the structure of the sets.  In
particular, we shall evolve the Julia sets of $F(a,s)$ along $a=0$ to
$a=-1$ along the real axis of $a$ as in Figure
\ref{fig:zjulia_evolve-}, and similarly but in opposite direction from
$a=0$ to $a=1$ along the real axis of $a$ as in Figure
\ref{fig:zjulia_evolve+}.

A numerical observation is that if sequence $F_n(a,s)$ in the Julia
set iterations converges, then $F_n(a,s)$ tends to a fixed point which
is a function $p_J(a)$ independent of $s$.  This translates to the
following conjecture:

\begin{conjecture}\quad\\
Given a fixed $a$, $\zeta(s) + a = s$ has either one or no finite
solution.
\end{conjecture}

\newpage

\newpage

\newpage

We shall now consider the Mandelbrot set of the Riemann zeta function.
Given the map $F(a,s) = \zeta(s) + a$ for which the initial iteration
value $s=s_0$ is a zero of the Riemann zeta function $\zeta(s)$, a
value of the complex parameter $a$ belongs to the Mandelbrot set of
$F$ if the iterated value $F_n(a,s_0)$ does not tend to complex
infinity in the limit the order of iteration, $n\to\infty$.

\bigskip

\begin{figure}[hbt]
\begin{center}
\begin{tabular}{rl}
$\bspace\bspace$\raisebox{2.5truein}{$\im(s)$}&
\includegraphics{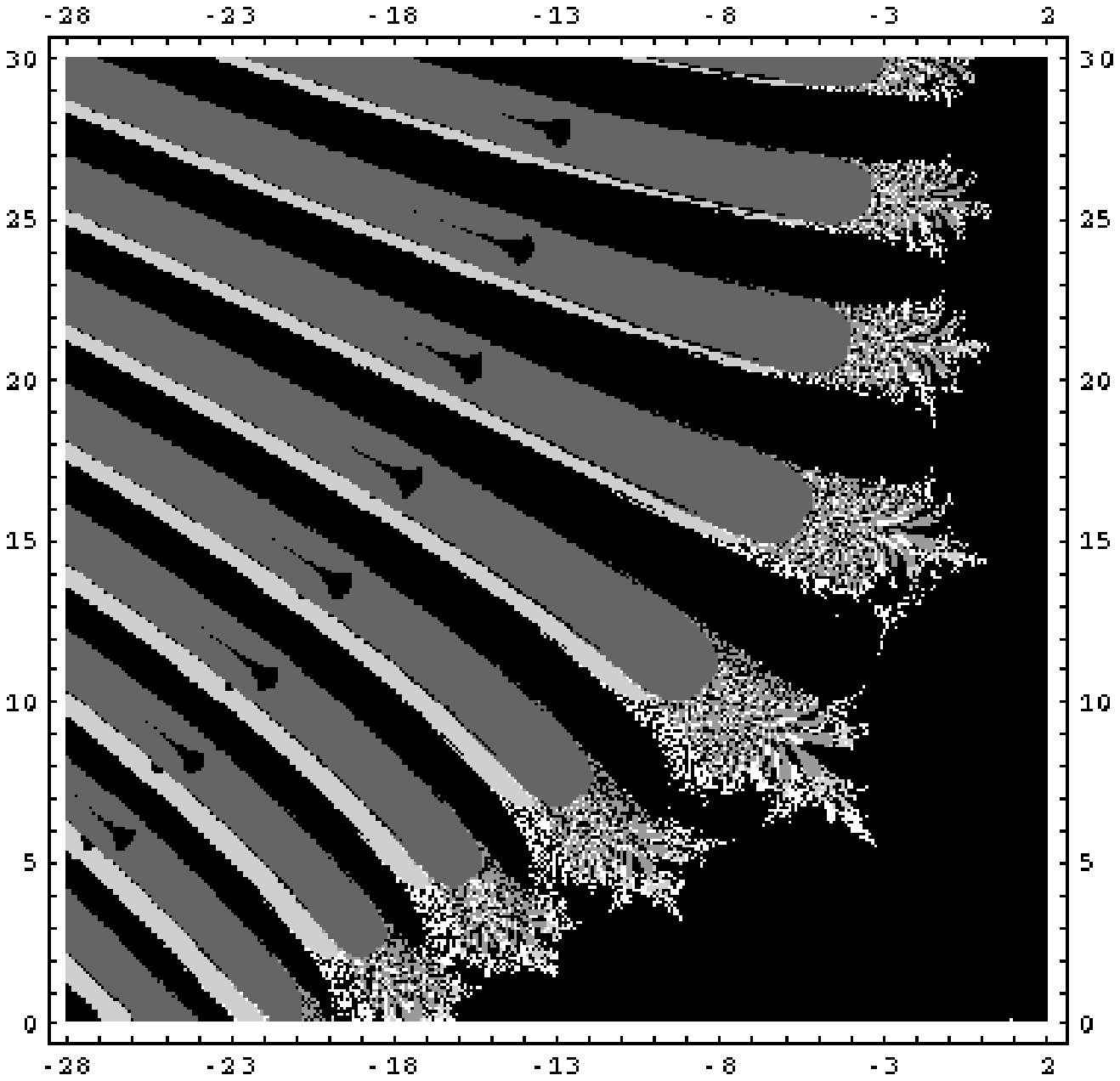}
\end{tabular}
$\re(s)$
\caption{Mandelbrot set of $F(a,s) = \zeta(s) + a$,}
\begin{tabular}{c}
for which $s$ is a zero of $\zeta(s)$,\\
$-28\le\re(a)\le 2\;\;$ and $\;\;0\le\im(a)\le 30\,$.
\end{tabular}
\label{fig:zmbrot}
\end{center}
\end{figure}

\begin{figure}[hbt]
\begin{center}
\begin{tabular}{rl}
$\bspace\bspace$\raisebox{2.5truein}{$\im(s)$}&
\includegraphics{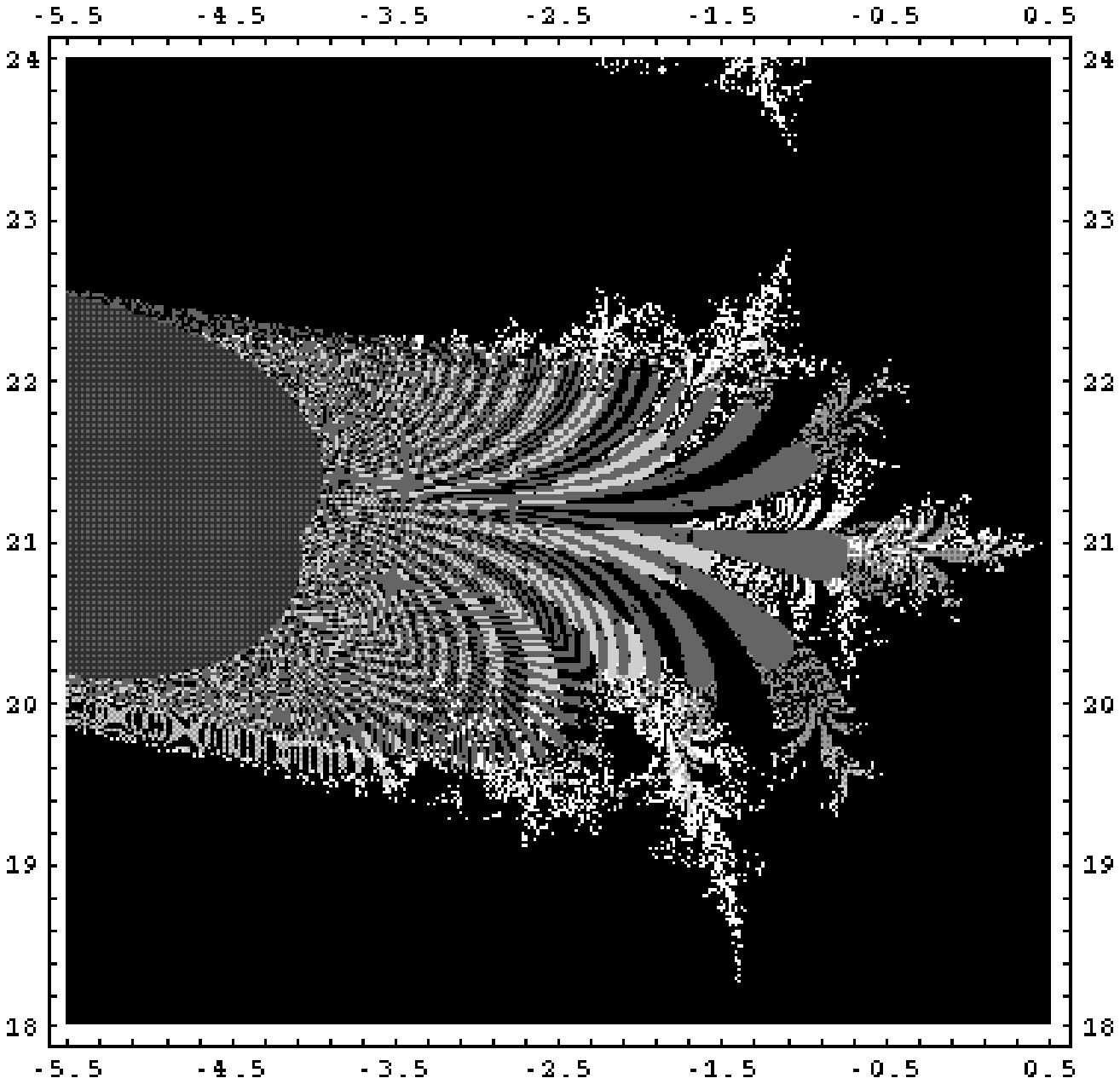}
\end{tabular}
$\re(s)$
\caption{Zooming in to reveal the self-similarity}
\begin{tabular}{c}
of the Mandelbrot set of $F(a,s) = \zeta(s) + a$\\
for which $s$ is a zero of $\zeta(s)$,\\
$-5.5\le\re(a)\le 0.5\;\;$ and $\;\;18\le\im(a)\le 24\,$.
\end{tabular}
\label{fig:zmbrot_zoom}
\end{center}
\end{figure}

\begin{figure}[hbt]
\begin{center}
\begin{tabular}{rl}
$\bspace\bspace$\raisebox{2.5truein}{$\im(s)$}&
\includegraphics{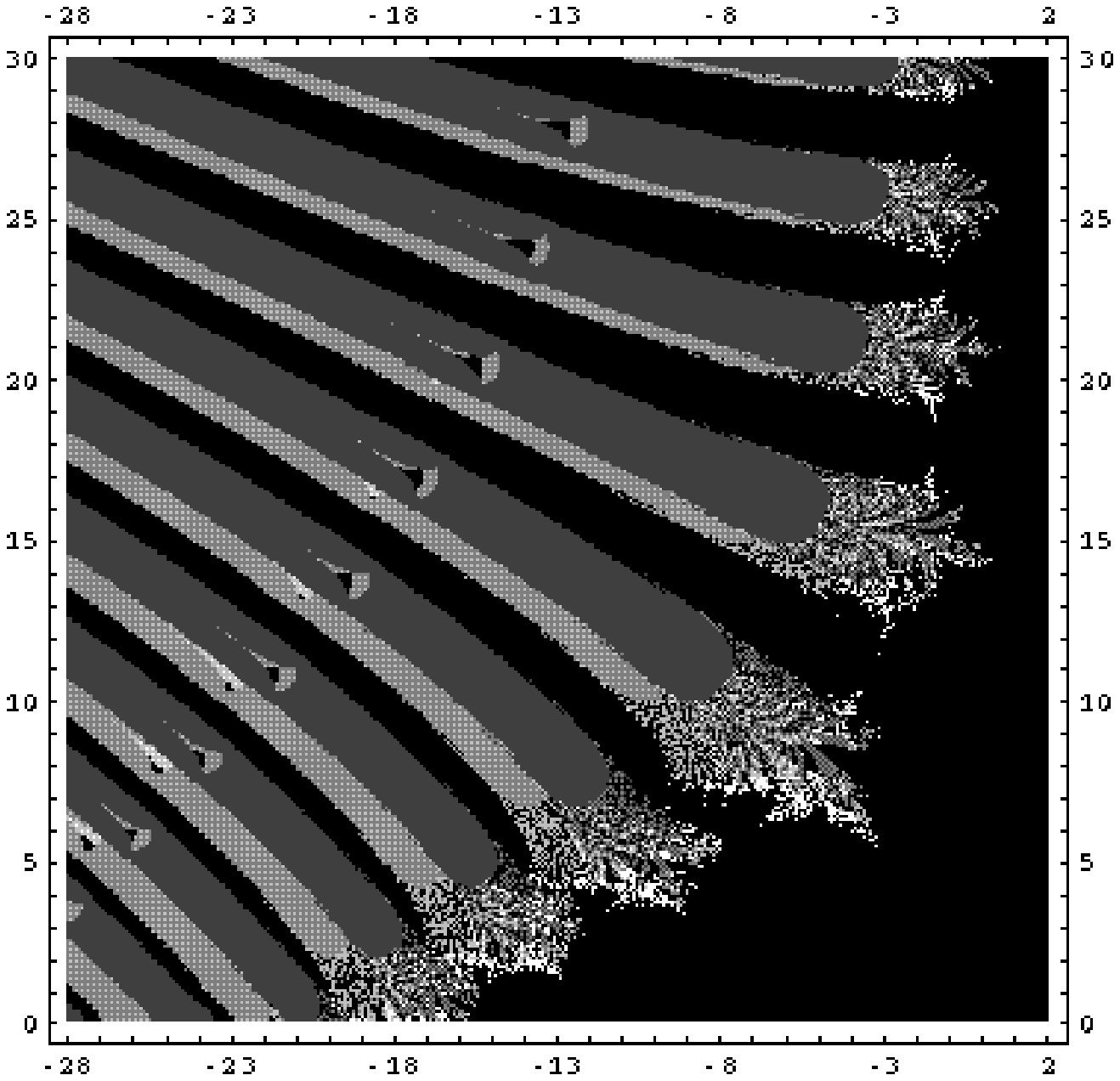}
\end{tabular}
$\re(s)$
\caption{Mandelbrot set of $F(a,s) = \zeta(s) + a\;$ for $\;s=0$,}
\begin{tabular}{c}
$-28\le\re(a)\le 2\;\;$ and $\;\;0\le\im(a)\le 30\,$.
\end{tabular}
\label{fig:zmbrot_s=0}
\end{center}
\end{figure}

The results of the computation of the Mandelbrot set of $F(a,s) =
\zeta(s) + a$ where the initial iteration value $s=s_0\in\{s :
\zeta(s) = 0$\}, are presented in Figures \ref{fig:zmbrot},
\ref{fig:zmbrot_zoom} and \ref{fig:zmbrot_s=0}.  This may also be the
very first attempt at such computation.

\bigskip

However, the following about the Mandelbrot set of the Riemann zeta
function was observed numerically and remains to be proved:

\begin{conjecture}\quad\\
For the map $F(a,s) = \zeta(s) + a$, if $a$ belongs to the
Mandelbrot set, then $F_n(a,s_0)$ in the Mandelbrot iterations tends
to a fixed point which is a function $p_M(a)$ independent of $s_0$, or
tends to a $n$-cycle where $n\in\Z^+$.
\end{conjecture}

In addition, if Theorem \ref{t:JuliaFatou} which applies to the
quadratic map also applies to $F(a,s) = \zeta(s) + a$, from Figure
\ref{fig:zmbrot}, we can see that $a=0$ belongs to the Mandelbrot set
of $F(a,s) = \zeta(s) + a$ for $s=0$, we deduce that the Julia set of
$\zeta(s)$ is connected.

\newpage

The black region represents the Mandelbrot set.  The regions of
different shades represent the various rates of amplitudes in
iteration $\to\infty$, with the lighter shades having slower rates,
and the darker shades, faster rates. There is a reflection symmetry
about the $\re(s)$-axis due to the complex conjugate property of an
analytic complex function.

The overall picture has the feature of a ``comb'' or ``bristles of a
brush'' with ``hairy ends''.  It is interesting to note that the
``bristle'' structures resemble those of the Julia sets for $|a|>>0$
in Figures \ref{fig:zjulia_evolve+} \ and \ref{fig:zjulia_evolve-} \ 
although the structures at the end of the ``bristles'' are different.

\newpage

The picture zooms onto the ``hairy end'' of a ``bristle'' only to
reveal more ``bristles'' structures with ``hairy ends'' features, and
ad infinitum.

\newpage

We can see that $a=0$ lies in the black region, and thus $a=0$ belongs to
Mandelbrot set of $F(a,s) = \zeta(s) + a\;$ for $\;s=0$.

We shall consider the more general case, the iterations of
fully-parametrized family of maps of the Riemann zeta function, in the
next paper.

\bigskip

\noindent{\bf \Large Appendix}

\appendix

\section{Examples of Approximate Self-Similarity and Scaling in Number Theory}

\subsection{Approximate self-similarity in the Riemann zeta 
function}

We shall define approximate self-similarity and show that the
following theorem due to Voronin implies that there are approximate
self-similarity in the images of the Riemann zeta function.

\begin{definition}[Approximate Self-Similarity to order $(\epsilon,r)$]
  An n-dimensional map $f$ has approximate self-similarity to order
  $(\epsilon,r)$ when given $\epsilon > 0$ and $r > 0$, there exist
  $X_a, X_b$ and $r_b$ such that $|f(X - X_a) - f(r_b (X - X_b))| <
  \epsilon$ for all $|X| < r$, where $X = (x_1,x_2,\dots,x_n)$.
\end{definition}

\begin{theorem}[{\bf Voronin}] \cite{voronin}\quad\\
\label{t:voronin}
Let $0<r<1/4$ and let $f(s)$ be a complex function analytic and
continuous for $|s|\le r$. If $f(s)\ne 0$, then for every $\epsilon>0$,
there exists a real number $T=T(\epsilon,f)$ such that
\begin{equation}
\max_{|s|\le r} \big| f(s)-\zeta(s+3/4+iT) \big| < \epsilon\;.
\label{e:voronin}
\end{equation}
\end{theorem}

\begin{theorem}
  There exists approximate self-similarity to order $(\epsilon,r)$ for
  any given $\epsilon>0,\; 0<r<1/4$ in the images of the Riemann zeta
  function.
\end{theorem}

{\bf Proof}

Choose $f(s) = \zeta(\rho\,e^{i\theta}\!s + a)$ with fixed
$\rho,\theta\in\R$ and fixed $a\in\C$ where $\rho>0,\; \theta\in[0,2
\pi),\; 0<|s|<1/4$, and choose $\epsilon$ to be arbitrarily small.
The expression $\rho\,e^{i\theta}\!s + a$ with fixed $\rho, \theta, a$
describes a disc centred at $a$ with radius $\rho/4$ and orientation
rotated at angle $\theta$.
  
By the inequality \ref{e:voronin} in Theorem \ref{t:voronin}, we have
self-similarities in the Riemann zeta function up to the order of
$\epsilon$ between the images of the discs at different centres,
scales and orientations (recentred by choosing a different fixed $a$,
rescaled by choosing a different fixed $\rho$, and rotated by choosing
a different fixed $\theta$), and the images of the discs along the
upper half of the critical strip at $s+3/4+iT$ with different values
of $T$.
\hfill\qed

\subsection{Scale-invariant equation for bounds in Goldbach conjecture}

\begin{conjecture}[{\bf Goldbach conjecture}]\quad\\
  Every even integer $2n\ge 4$ can be written as a sum of two primes.
\end{conjecture}

\begin{definition}\quad\\
  Define $G(2n)$ as the number of distinct decompositions of the even
  integer $2n$ into a sum of two primes ($3+5$ and $5+3$ are counted
  as identical decomposition).
  
  Define $G_L(x)$ and $G_U(x)$ as the lower and upper monotonic
  bounding curves respectively of $G(2n)$ such that
  $$\frac{d G_L(x)}{dx} > 0 \;\;\mbox{ and }\;\; \frac{d G_U(x)}{dx} \;\mbox{
    for all } x\;,$$
  and $G_L(2n) \le G(2n) \le G_U(2n)$ for all even integer $2n\ge 4$.
\end{definition}

Thus, Goldbach conjecture states that $G_L(2n)\ge 1$ for all even
integer $2n\ge 4$.  However, the plots of $G(2n)$ in Figure
\ref{fig:goldbach} seem to show that the shapes of the bounding curves
are invariant under scaling.  This observation leads us to suggest a
new conjecture:

\begin{figure}[hbt]
\begin{center}
\begin{tabular}{rccc}
$\bspace\bspace\bspace$\raisebox{0.5truein}{$G(n)$}&
\includegraphics[height=1truein]{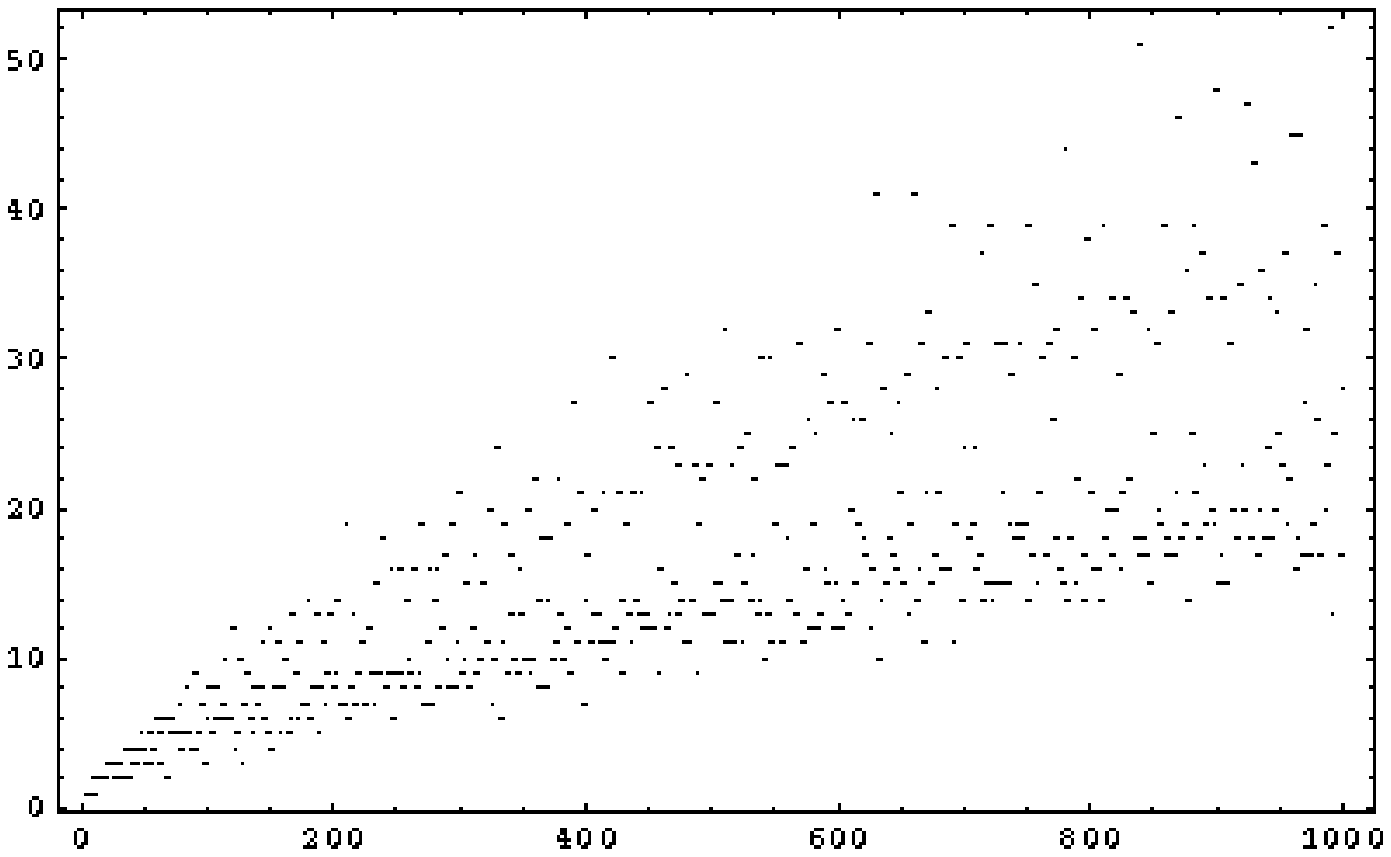}&
\includegraphics[height=1truein]{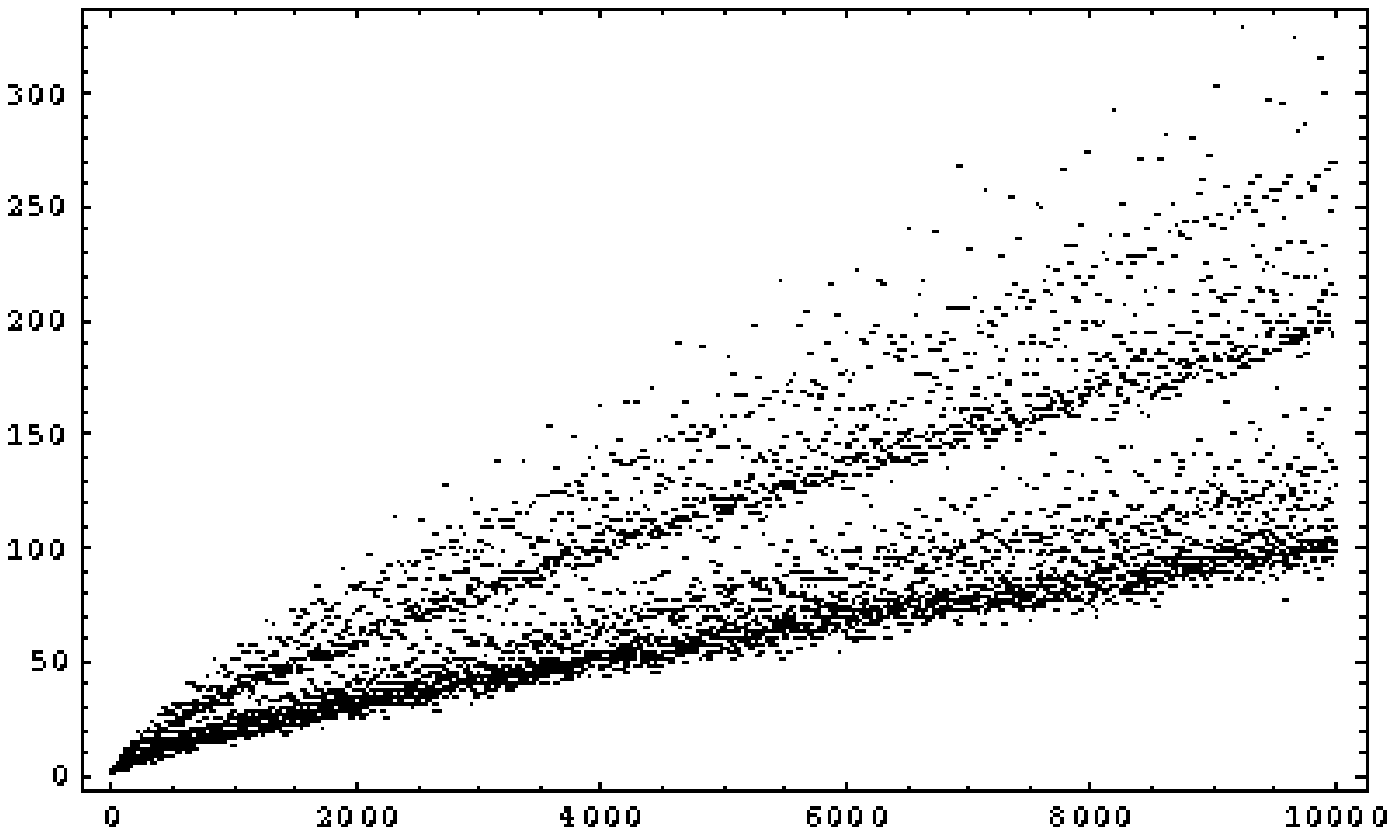}&
\includegraphics[height=1truein]{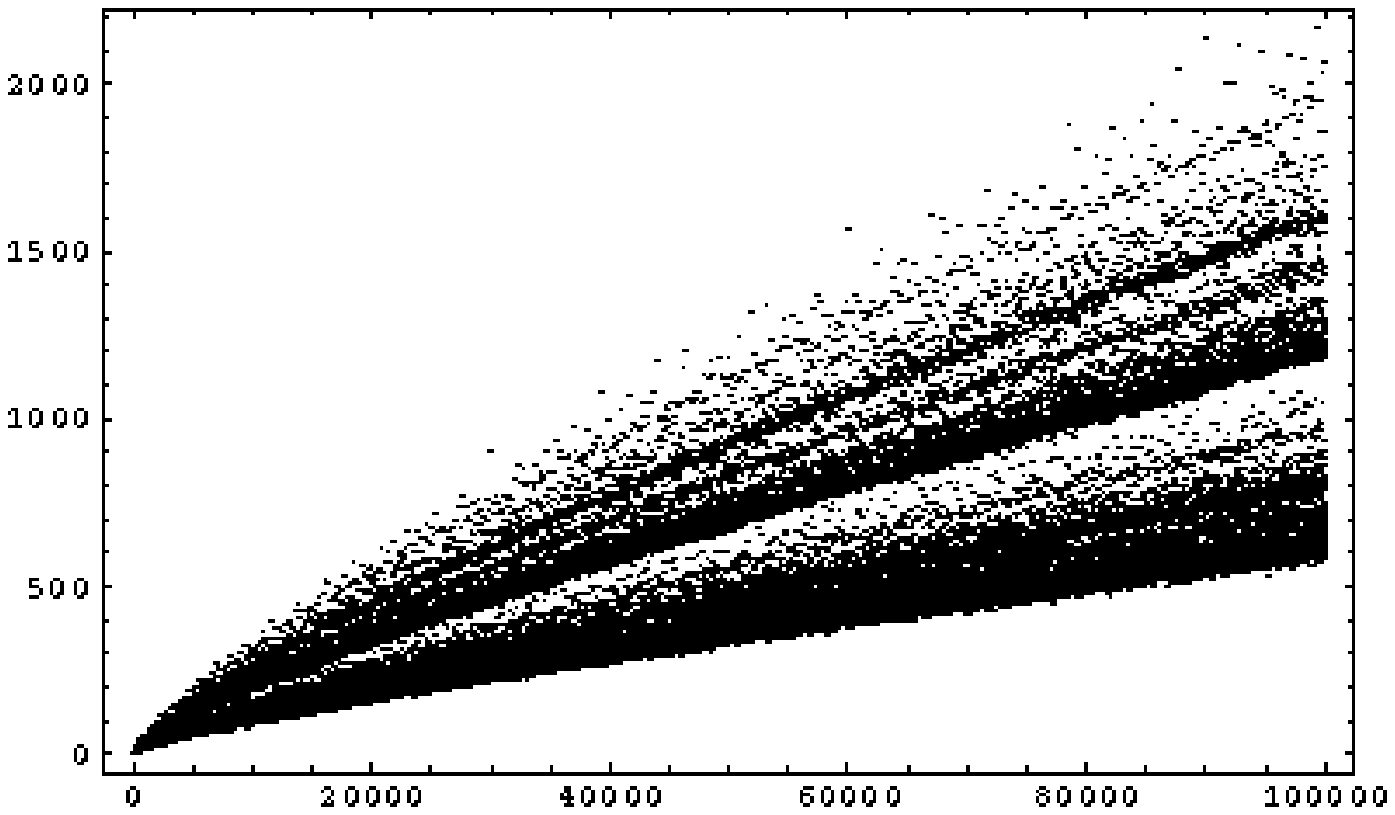}\\
&$n$&$n$&$n$
\end{tabular}
\caption{Scale invariance in the bounding curves of $G(2n)$ and the 
  approximate self-similar pattern in $G(2n)$}
\label{fig:goldbach}
\end{center}
\end{figure}

\begin{conjecture}[{\bf Scale-invariant equation for bounds in Goldbach
    conjecture}]
$G_L(x)$ satisfies the scale invariance equation
\begin{eqnarray}
&&\quad \frac{G_L(ax)}{G_L(a)} \;=\; \frac{G_L(bx)}{G_L(b)}
\quad(\mbox{for all real }\, a,b\ne 0)\nn\\
&\Rightarrow& \frac{G_L(a)}{G_L(b)}\;G_L(bx/a) \;=\; G_L(x)\;,
\label{e:invariance}
\end{eqnarray}
and similarly for $G_U(x)$.
\end{conjecture}

Although $f(ax) = f(a)\,f(x) \;$ (for all $a$) satisfies
(\ref{e:invariance}), the form of $G_L(x)$ is not in the form of
$f(x)$.  Since the simple power law $y(x) =
\beta\,x^\lambda,\;\beta,\lambda\in\R$ is in the form of $f(x)$, we
thus deduce that $G_L(x)$ is not a simple power law.

\begin{figure}[hbt]
\begin{center}
\begin{tabular}{rl}
$\bspace\bspace$\raisebox{0.8truein}{$G(n)$}&
\includegraphics[height=1.8truein]{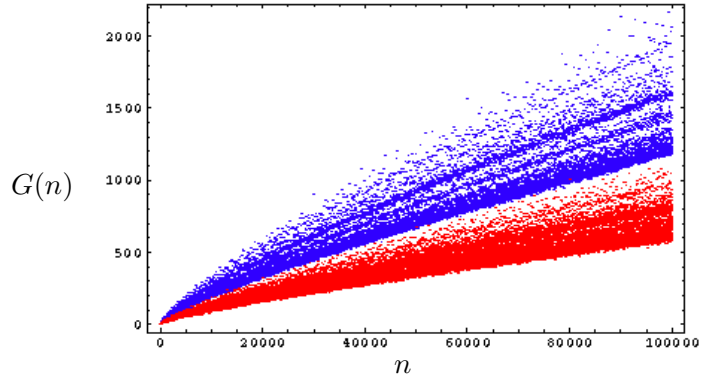}
\end{tabular}\\
$n$
\caption{The subset of $\{G(2n)\}$ with $n$ being a multiple of $3$,
  $\{G(2(3n))\}$, is labelled by blue dots. The subset which is not,
  $\{G(2(3n+1))\} \cap \{G(2(3n+2))\}$, is labelled by red dots.}
\label{fig:prime3}
\end{center}
\end{figure}

\begin{figure}[hbt]
\begin{center}
\begin{tabular}{rl}
$\bspace\bspace$\raisebox{0.8truein}{$G(n)$}&
\includegraphics[height=1.8truein]{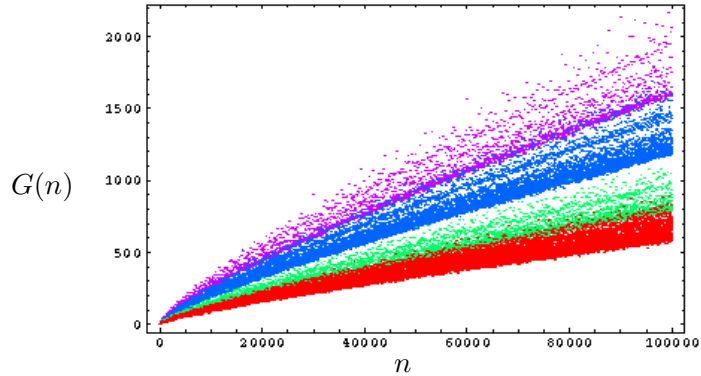}
\end{tabular}\\
$n$
\caption{$\{G(2n)\;|\;n\equiv 0$ mod $3,\; n\equiv 0$ mod $5\}$ is
  labelled by magenta dots, $\{G(2n)\,|\,n\equiv 0$ mod $3,\;
  n\not\equiv 0$ mod $5\}$ by blue dots, $\{G(2n)\,|\,n\not\equiv 0$
  mod $3,\; n\equiv 0$ mod $5\}$ by green dots,
  $\{G(2n)\,|\,n\not\equiv 0$ mod $3,\; n\not\equiv 0$ mod $5\}$ by
  red dots.}
\label{fig:prime3and5}
\end{center}
\end{figure}

The patterns in $G(2n)$ as illustrated by Figures \ref{fig:prime3} and
\ref{fig:prime3and5} are due to the distribution of prime numbers.
Hardy and Littlewood \cite{hardygoldbach} had derived an estimate for
$G(2n)$ which accounts for the patterns,
$$G(2n) \sim \frac{2\,n\,C}{(\log n)^2} \prod_{p>2,\; p|n} \frac{p-1}{p-2}$$
where
$p$ runs over all prime numbers, $p|n$ denotes that the product is
taken only if $n$ is divisible by $p$, and $C$ is a constant,
$$C = \prod_{p>2} \left(1 - \frac{1}{(p-1)^2}\right) = 0.66016\dots\;.$$

However, the scale-invariant property of the bounding curves $G_L(x)$
and $G_U(x)$ of $G(2n)$ as expressed in (\ref{e:invariance}) is a new
observation.

\subsection{Open Problem}

\begin{enumerate}
  
\item Derive the exact expressions for the bounding curves $G_L(x)$
  and $G_U(x)$.

\end{enumerate}

We shall attempt to derive $G_L(x)$ and $G_U(x)$ in the next paper.

\noindent{\bf Acknowledgement}

Thanks to Michael Rubinstein, Keith Briggs and Gove Effinger for
helpful comments.

\end{document}